\documentclass{article}%
\usepackage{amsmath}
\usepackage{graphicx}
\usepackage{amsfonts}
\usepackage{amssymb}%
\begin{document}

\title{The Conductance of a Perfect Thin Film with Diffuse Surface Scattering.}
\date{\today}
\author{Gerd Bergmann\\Physics Department, Univ.South.California\\Los Angeles, CA 90089-0484, USA}
\maketitle

\begin{abstract}
The conductance of thin films with diffusive surface scattering was solved
semi-classically by Fuchs and Sondheimer. However, when the intrinsic electron
mean free path is very large or infinite their conductance diverges. In this
letter a simple diffraction picture is presented. It yields a conductance
which corresponds to a limiting mean free path of $a^{2}k_{F}/(2\pi)$ where
$a$ is the film thickness.


\end{abstract}

The conductivity of a thin film generally decreases considerably when the film
thickness is reduced. The main reason is that the surfaces of the film scatter
the conduction electrons partially or completely diffusively and reduce their
effective mean free path. This phenomenon was first treated by Fuchs
\cite{F31} and later generalized by Sondheimer \cite{S36}. The result of their
semi-classical calculation for the conductance $G$ of a square shaped thin
film (width equal to length) of thickness $a$ with completely diffusive
scattering at the surfaces is
\begin{equation}
G_{\square}=\frac{ne^{2}al_{0}}{mv_{F}}\left[  1-\frac{3l_{0}}{2a}\int_{0}%
^{1}\left(  1-u^{2}\right)  u\left\{  1-\exp\left(  -\frac{a}{l_{0}u}\right)
\right\}  du\right]
\end{equation}
where $n$ is the density, $l_{0}$ the bulk mean free path and $v_{F}$ the
Fermi velocity of the electrons. $u=\sin\alpha,$ where $\alpha$ is the angle
between the velocity vector of the electrons and the film plane. Chambers
\cite{C17} interpreted these results by introducing an $\mathbf{\ r}$- and
$\mathbf{k}$-dependent vector mean free path $\mathbf{l}_{\mathbf{r,k}}$
$=\int_{-t_{0}}^{0}\mathbf{v}\left(  t^{\prime}\right)  e^{t^{\prime}/\tau
}dt^{\prime}$ . The vector mean free path in a film is reduced by the
diffusive scattering at the surface (at the earlier time $-t_{0})$ and depends
therefore on the position in the film and the z-component of the Fermi
velocity. While the conduction electron travels along the vector mean free
path it accumulates momentum and energy. The energy is equal to $\left(
-e\right)  \mathbf{E\cdot l}_{\mathbf{k}}$ and yields a correction
$f_{1}\left(  \mathbf{r,k}\right)  =\left(  -\frac{df_{0}}{d\varepsilon
}\right)  \left(  -e\right)  \mathbf{E\cdot l}_{\mathbf{r,k}}$ to the
distribution function $f\left(  \mathbf{r,k}\right)  $ $=f_{0}+f_{1}\left(
\mathbf{r,k}\right)  $ where $f_{0}$ is the equilibrium Fermi function.

The Fuchs-Sondheimer conductance of a thin film diverges when the intrinsic
mean free path becomes infinite. For very large intrinsic mean free path the
conductance of a thin film becomes \cite{S36}
\begin{equation}
G_{\square}=\frac{ne^{2}}{mv_{F}}a\ast\frac{3a}{4}\ln\frac{l_{0}}{a}=\left(
\frac{a}{\lambda_{F}}\right)  ^{2}\frac{e^{2}}{\hbar}\ln\frac{l_{0}}{a}
\label{Slmfp}%
\end{equation}
where $\lambda_{F}=2\pi/k_{F}$ is the Fermi wave length.

The divergence is due to the electrons which travel almost parallel to the
surface, so that their effective mean free path becomes very large. An
electron with $v_{z}=v_{F}\sin\alpha$ travels the distance $av_{F}%
/v_{z}=a/\sin\alpha\thickapprox a/\alpha$ from one surface to the opposite one
if there are no internal collisions. For a given $v_{z}$ the distance traveled
since the last collision increases linearly with the distance from the surface
of last collision from $0$ to $av_{F}/v_{z}$. Therefore the average mean free
path for electrons with a given $v_{z}$ or a given angle $\alpha$ is half the
distance between two surface collisions $l_{ef}=a/\left(  2\alpha\right)  $.
This $l_{ef}$ diverges for $\alpha\rightarrow0$.

It has been pointed out in the past that this divergence is a classical effect
which should disappear if one treats the problem quantum theoretically.
Tesanovic et al. \cite{M44} performed such a calculation. They considered a
thin film whose thickness is modulated (modeling a finite surface roughness)
and transformed the surface modulation into a scattering potential. In their
evaluation they expanded the wave function in the basis of standing waves in
the z-direction perpendicular to the film plane, $\psi\varpropto\sin\left(
k_{z}z\right)  $. This yields a quantization of $k_{z}=n_{z}\pi/a$,
$n_{z}=1,2,..$. Comparison with experimental data yielded a reasonable
modulation amplitude for the film thickness.

For a completely diffusive scattering at the surface the average (vector) mean
free path in z-direction is equal to half the film thickness $a/2$. This
yields an uncertainty of $\Delta k_{z}$ which is about $2/a$, i.e., of the
order of the separation of the $k_{z}$-values. Therefore this approach might
not be well suited for a surface with completely diffusive scattering.

In this letter we give (i) a semi-quantitative evaluation of the scattering in
the Feynman path integral approach and (ii) treat the problem as an electron
diffraction by the film boundaries. Both treatments yield essentially the same
quantitative results that the diffusive surface scattering acts as if the
conduction electrons experience an effective mean free path of $l_{0}%
=a^{2}/\left(  2\lambda_{F}\right)  $ ($\lambda_{F}$ =2$\pi/k_{F}$= Fermi wave length).

The Kubo formalism provides a tool to calculate the conductance for an
arbitrary geometry. In the real space representation the Kubo formula uses the
Green function $G_{\omega}\left(  \mathbf{r,r}^{\prime}\right)  $ for the
propagator from $\mathbf{r}$ to $\mathbf{r}^{\prime}$. If one uses the
following form of the real space Green function
\begin{equation}
G_{\omega}\left(  \mathbf{r-r}^{\prime}\right)  =\frac{m}{2\pi\hbar}\frac
{1}{\left\vert \mathbf{r-r}^{\prime}\right\vert }\left[  \exp\left(
ik_{F}\left\vert \mathbf{r-r}^{\prime}\right\vert \text{sgn}\left(
\omega\right)  -\frac{\left\vert \omega\right\vert }{v_{F}}\left\vert
\mathbf{r-r}^{\prime}\right\vert \right)  \right]  \label{GF}%
\end{equation}
then one can reproduce the vector mean free path result for the conductance
(which is identical with the Fuchs-Sondheimer formula).

$%
\raisebox{-0pt}{\parbox[b]{4.3441in}{\begin{center}
\includegraphics[
height=0.9297in,
width=4.3441in
]%
{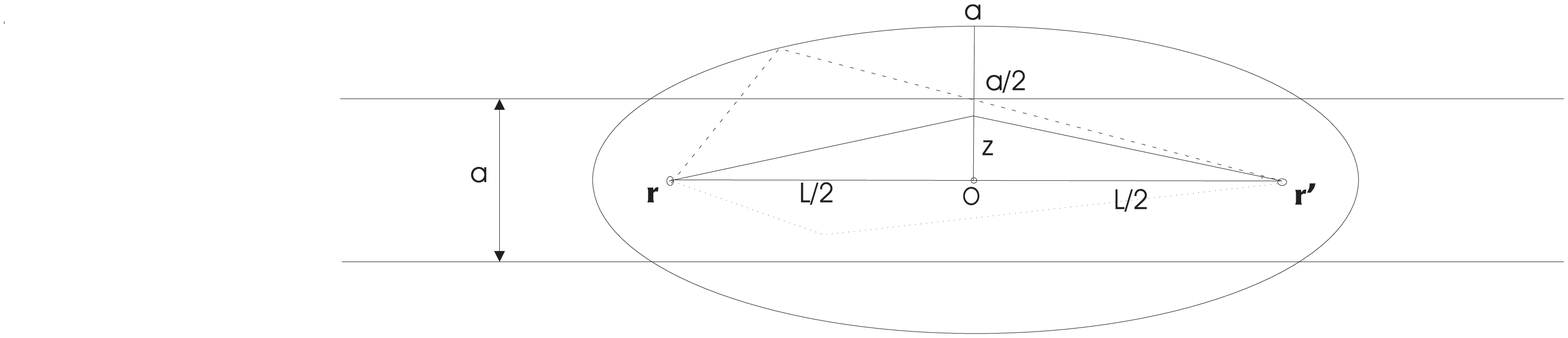}%
\\
Fig.1: Different Feynman paths from $\textbf{r}$ to $\textbf{r}^\prime$ in a
thin film via an intermediate point $\textbf{r}^\prime\prime$. All these paths
inside the ellipse are coherent with the direct path. The diffusive scattering
at the film surfaces eliminates many of the coherent paths.
\end{center}}}%
$

In a thin film with diffuse surface scattering the Green function can no
longer be expressed by equation (\ref{GF}). The reason can be understood in
terms of Feynman's path integral. A (quantum theoretical) electron does not
propagate along a straight line with constant velocity $v_{F}$ from
$\mathbf{r}$ to $\mathbf{r}^{\prime}$ but uses all possible paths from
$\mathbf{r}$ to $\mathbf{r}^{\prime}$. Of course, most of these paths are not
important because their contribution averages to zero due to their phase
$\phi=\exp\left[  \frac{i}{\hbar}\int_{0}^{T}L\left(  \mathbf{r,}%
\overset{\circ}{\mathbf{r}}\right)  dt\right]  $, ($L\left(  \mathbf{r,}%
\overset{\circ}{\mathbf{r}}\right)  =\frac{m}{2}\left(  \overset{\circ
}{\mathbf{r}}\right)  ^{2}$ is the Langrangian). However, paths which are
sufficiently close to a straight line contribute to the amplitude since one
has to sum over all paths.

In Fig.1 a thin film is drawn with the position $\mathbf{r=}\left(
-L/2,0,0\right)  $ in the center of the film and the position $\mathbf{r}%
^{\prime}=\left(  L/2,0,0\right)  $ at a distance $L$ from $\mathbf{r}$. The
film plane is perpendicular to the z-direction and the electric field is
applied in x-direction. The traveling time for an electron with Fermi velocity
is $T=L/v_{F}$ and the phase of the direct path is $\phi_{0}=\left(
m/2\right)  v_{F}^{2}T/\hbar$. If we consider a very simple alternative path
in which the electron moves from $\mathbf{r=}\left(  0,0,0\right)  $ to
$\mathbf{r}^{\prime\prime}=\left(  0,0,z\right)  $ in the time interval $T/2$
and then from $\mathbf{r}^{\prime\prime}$ to $\mathbf{r}^{\prime}$ its average
velocity is $v^{\prime}=v_{F}\sqrt{1+\left(  z/2L\right)  ^{2}}$. This yields
for the phase of this path $\phi=\left(  m/2\right)  v_{F}^{2}\left(
1+\left(  z/2L\right)  ^{2}\right)  T/\hbar$ $=\phi_{0}+k_{F}z^{2}/\left(
8L\right)  .$ As long as $\Delta\phi=k_{F}z^{2}/\left(  8L\right)  $%
$<$%
$\pi/2$ this path and similar ones contribute to the amplitude at
$\mathbf{r}^{\prime}$.

In Fig.1 we have chosen $L=k_{F}a^{2}/\left(  4\pi\right)  $ and drawn an
ellipse with focal points at $\left(  \pm L/2,0,0\right)  $ and the short semi
axis equal to the film thickness $a$. All paths $\mathbf{r\rightarrow
r}^{\prime\prime}\rightarrow\mathbf{r}^{\prime}$ where $\mathbf{r}%
^{\prime\prime}$ lies within the ellipse have a phase difference which is less
than $\pi/2$ with respect to the direct path, all paths where $\mathbf{r}%
^{\prime\prime}$ lies outside the ellipses have a phase difference larger than
$\pi/2$. For this situation a large fraction of the constructive paths
intersects the film boundary and is eliminated because of the diffuse
scattering at the surface. We can roughly say that the fraction of electrons
which reach the point $\mathbf{r}^{\prime}$ without phase randomization is
reduced by a factor of two. That means that the mean free path of the
electrons parallel to the film is of the order of $k_{F}a^{2}/\left(
4\pi\right)  $. So their contribution to the conductance is finite. One
obtains roughly the same reduction when the z-components of $\mathbf{r}$ and
$\mathbf{r}^{\prime}$ are varied between $\pm a/2$. This means that all paths
within the angle $\sin\alpha$=$a/L=4\pi/\left(  ak_{F}\right)  =2\lambda
_{F}/a$ have a finite mean free path of about $a^{2}/\left(  2\lambda
_{F}\right)  $.

The quantitative evaluation of the Feynman path integral is a rather complex
task, in particular the normalization of the sum or integral over the
different paths. Therefore we turn to a different and quite simple approach
which I call the diffraction model.

If we cut the film at a plane parallel to the y-z-plane then the cross section
of the film represents a slit with the width $a$. This is shown in Fig.2.
Through this slit the conduction electron waves propagate with the wave length
$\lambda_{F}=\frac{2\pi}{k_{F}}$. Therefore the electron waves experience
diffraction. The result of the diffraction is that the wave spreads out. The
angle for the first diffraction minimum is
\[
\sin\alpha_{d}=\frac{\lambda_{F}}{a}=\frac{2\pi}{k_{F}a}%
\]%
\[%
\begin{tabular}
[c]{l}%
{\includegraphics[
height=2.7347in,
width=4.3769in
]%
{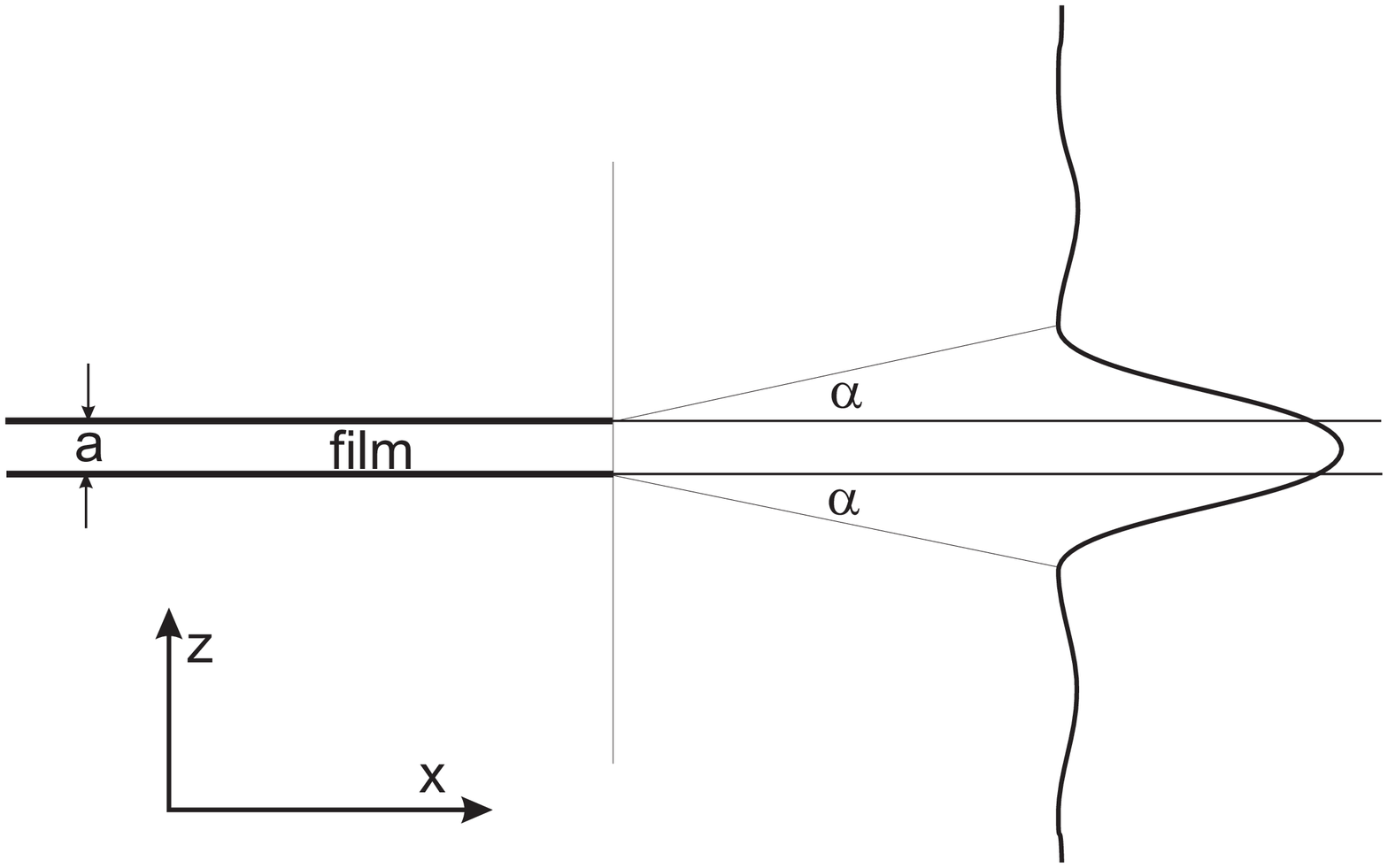}%
}
\\
Fig.2: The film is cut perpendicular to show the spread of the \\
electron wave function due to diffraction by a single slit. The \\
curve on the right gives the single slit diffraction intensity. \\
The part of the electron wave outside of the film (dashed lines) \\
will be diffusively scattered. Therefore the electron wave is \\
dampened.
\end{tabular}
\ \ \
\]

Now we consider an electron which travels parallel to the film. Then the slit
experiment tells us that at the distance $dx=v_{F}dt$ after the slit the beam
of width $a$ has broadened to $\left(  a+2\alpha_{d}dx\right)  =\left(
a+2\alpha_{d}v_{F}dt\right)  $. Since the fraction of the wave outside of $a$
is scattered at the surface the remaining intensity $I$ of the electron within
the film reduces to $\left(  I+dI\right)  $, $dI<0$%
\begin{align*}
I+dI  &  =I\frac{a}{a+2\alpha_{d}dx}\thickapprox I\left(  1-\frac{2\alpha_{d}%
}{a}dx\right) \\
dI  &  =-I\frac{2\alpha_{d}}{a}dx=-I\frac{2\lambda_{F}}{a^{2}}dx=-I\frac
{2\lambda_{F}v_{F}}{a^{2}}dt\\
I  &  =I_{0}\exp\left(  -\frac{2\alpha_{d}x}{a}\right)  =I_{0}\exp\left(
-\frac{2\alpha_{d}v_{F}t}{a}\right)
\end{align*}
As a consequence the mean free path of the electron which travels parallel to
the film, is $l_{d}=\frac{a}{2\alpha_{d}}=\frac{a^{2}}{2\lambda_{F}}$ or the
relaxation time is $\tau=\frac{a}{2\alpha_{d}v_{F}}=\frac{a^{2}m}{4\pi\hbar}$.
Since the diffraction of the electron wave occurs on both film surfaces within
an angle $\alpha_{d}$ all electrons which travel almost parallel to the
x-y-plane within the angle $\alpha_{d}$ have the same mean free path
$l_{d}=a^{2}/\left(  2\lambda_{F}\right)  $. Actually $l_{d}$ is identical
with half the distance it takes an electron to travel from one surface to the
opposite one at the angle $\alpha_{d}$. For larger angles $\alpha$ between the
wave vector and the film plane the diffraction has little effect, even a
classical electron hits the surface and the mean free (averaged over the
position in the film is $a/(2\sin\alpha)$ which is shorter then $a/\left(
2\alpha_{d}\right)  $.

We obtain the quantum theoretical conductance of a thin film with diffuse
surface scattering but infinite intrinsic mean free path by dividing the
integration over $du=d\left(  \sin\alpha\right)  $ into two regions; for
$\left\vert \alpha\right\vert >\left\vert \alpha_{d}\right\vert $ we use an
infinite mean free path and for $\left\vert \alpha\right\vert <\left\vert
\alpha_{d}\right\vert $ the finite mean free path of $a^{2}/\left(
2\lambda_{F}\right)  $. We find for $G_{\infty}$
\begin{equation}
G_{\infty}=\frac{e^{2}}{\hbar}\left[  \left(  \left(  \frac{a}{\lambda_{F}%
}\right)  ^{2}\left(  \ln\frac{a}{\lambda_{F}}+\frac{1}{2}\right)  +\frac
{1}{6}\right)  \right]  \label{Ilmfp}%
\end{equation}
This result has logarithmic accuracy only. Therefore we ignore the
non-logarithmic terms and obtain as our final result%
\begin{equation}
G_{\infty}\cong\frac{e^{2}}{\hbar}\left(  \frac{a}{\lambda_{F}}\right)
^{2}\left(  \ln\frac{a}{\lambda_{F}}\right)  \label{reslt}%
\end{equation}
The conductance depends only on the ratio $a/\lambda_{F}.$

Obviously the conductance does not diverge in contrast to the semi-classical
solution by Fuchs and Sondheimer. Because of the wave character of the
electrons the diffuse surface scattering introduces an effective mean free
path. One can incorporate this result in the semi-classical solution by
replacing the intrinsic mean free path in equation (\ref{Slmfp}) by
$a^{2}/\lambda_{F}$.

Acknowledgment: The research was supported by NSF Grant No. DMR-0124422.
\[
\]

\end{document}